\documentclass[a4paper,11pt]{article}

\usepackage{jheppub}
\usepackage{graphicx}
\usepackage{amsmath}
\usepackage{xspace}

\usepackage{hyperref}
\usepackage{color}

\newcommand{\be}{\begin{equation}}
\newcommand{\ee}{\end{equation}}
\newcommand{\bea}{\begin{eqnarray}}
\newcommand{\eea}{\end{eqnarray}}

\newcommand{\MS}{\overline{\rm MS}}

\newcommand{\nn}{\nonumber}

\def\lQ{\Lambda_{\mathrm{QCD}}}

\def\al{\alpha}

\def\siml{{\ \lower-1.2pt\vbox{\hbox{\rlap{$<$}\lower6pt\vbox{\hbox{$\sim$}}}}\ }}

\def\siml{{\ \lower-1.2pt\vbox{\hbox{\rlap{$<$}\lower6pt\vbox{\hbox{$\sim$}}}}\ }}

\def\bfsigma{\mbox{\boldmath $\sigma$}}

\title
{Model-independent determination of the two-photon exchange contribution to hyperfine splitting in muonic hydrogen} 

\author{Clara Peset,}
\author{Antonio Pineda,}

\emailAdd{peset@ifae.es}
\emailAdd{pineda@ifae.es}

\affiliation{Grup de F\'\i sica Te\`orica, Dept. F\'\i sica and IFAE-BIST, Universitat Aut\`onoma de Barcelona,\\ 
E-08193 Bellaterra (Barcelona), Spain}

\abstract{
We obtain a model-independent prediction for the two-photon exchange contribution to the hyperfine splitting in muonic hydrogen. 
We use the relation of the Wilson coefficients of the spin-dependent dimension-six four-fermion operator of NRQED applied to the electron-proton and 
to the muon-proton sectors. Their difference can be reliably computed using chiral perturbation theory, whereas the Wilson coefficient of the electron-proton sector can be determined from the hyperfine splitting in hydrogen. This allows us to give a precise model-independent determination of the Wilson coefficient for the
muon-proton sector, and consequently of the two-photon exchange contribution to the hyperfine splitting in muonic hydrogen, 
which reads $\delta \bar E_{p\mu,\rm HF}^{\rm TPE}(nS)=-\frac{1}{n^3}1.161(20)$ meV. Together with the associated QED analysis, we obtain a prediction for the 
hyperfine splitting in muonic hydrogen that reads $E^{\rm th}_{p\mu,\rm HF}(1S)=182.623(27)$ meV and $E^{\rm th}_{p\mu,\rm HF}(2S)=22.8123(33)$ meV. The error is dominated by the two-photon exchange contribution.
}

\begin{document}

\maketitle

\section{Hyperfine splitting in Hydrogen}

The hyperfine splitting of the ground state of hydrogen is one of the most accurate measurements made by mankind~\cite{exp,exp2,exp3,exp4,exp5,exp6,exp7,exp8}. One has  
\be
E_{\rm hyd,HF}^{\rm exp}(1S)=1420.405751768(1)\;{\rm MHz}
\,,
\ee
where we take the average from Ref.~\cite{Karshenboim:2005iy}.

Since then, there has been a continuous effort to derive such number from theory.   
The first five digits of this number can be reproduced by the theory of an infinitely massive proton (except for the  
$1/m_p$ prefactor of the Fermi term incorporating the anomalous magnetic moment of the proton) and a non-relativistic lepton, systematically incorporating the relativistic corrections of the latter. A summary of these pure QED computations can be found in Refs.~\cite{BY,Karshenboim:2005iy,Eides:2000xc}. Particularly detailed is the account of Ref.~\cite{Eides:2000xc}, which we take for reference. Such computation has reached (partial) ${\cal O}(m_e\al^8)$ precision and reads (compared to the notation of Ref.~\cite{Peset:2015zga} we set $Z_e=1$ and $Z=Z_p$ (=1 for numerics))
\bea
\nn
&&
\delta E_{\rm hyd,HF}^{\rm QED}(1S)=
\frac{8m_r^3\alpha(Z\alpha)^3}{3m_pm_e} c_F^{(p)}c_F^{(e)}
+
\frac{8m_r^3\alpha(Z\alpha)^3}{3m_pm_e} c_F^{(p)}Z\al^2
\left\{ \frac{3}{2}Z +\left[\ln 2-\frac{5}{2}\right]
\right.
\\
&+& \frac{Z\alpha}{\pi}\left[-\frac{2}{3} \ln ^2\frac{1}{(\alpha  Z)^2}-\left(\frac{8 \ln 2}{3}-\frac{281}{360}\right) \ln \frac{1}{(\alpha  Z)^2}+\frac{34}{225}+17.1227(11)-\frac{8 \ln 2}{15}\right]\nn\\
&+&
\frac{\alpha}{\pi}\left[\frac{\pi ^2}{9}-\frac{38 \pi }{15}+\frac{91639}{37800}-1.456(1)
-\frac{4}{3}\left(\ln\frac{1+\sqrt{5}}{2}+\frac{5}{3}\sqrt{5}\right)  \ln \frac{1+\sqrt{5}}{2} +\frac{608 \ln 2}{45}\right]
\nn\\
&+&
\frac{Z\alpha^2}{\pi^2}\left[10(2.5)-\frac{1}{3} \ln ^2\frac{1}{(\alpha  Z)^2}+\left(\frac{27 \zeta (3)}{16}+\frac{25187}{8640}+\pi ^2 \left(\frac{133}{432}-\frac{9 \ln 2}{8}\right)-\frac{4 \ln 2}{3}\right) \ln\frac{1}{(\alpha  Z)^2}\right]
\nn\\
&+& Z^2\alpha^2\left[\left(\frac{5 \ln 2}{2}-\frac{547}{96}\right) \ln 
\frac{1}{(\alpha  Z)^2}-3.82(63)\right]
\nn\\
&+& Z^3\alpha^2\left.\left[\frac{17}{8}\right]+{\cal O}(Z^0\al^2)\right\}
\,,
\label{QED}\eea
where $m_r=m_pm_e/(m_p+m_e)$.  $c_F^{(p)}\equiv Z+\kappa_p=2.792847356$ 
and $c_F^{(e)}=1.00115965$
 are the magnetic moments of the proton and electron respectively, which we take exactly, i.e. they include the ${\cal O}(\al)$ corrections. 
 Besides those, there are pure QED recoil corrections of ${\cal O}\left(m_e \al^6 \frac{m_e}{m_p}\right)$, 
 computed in Ref.~\cite{BY} (we take $Z=1$ in this expression):
 \bea
 \label{recoil}
 \delta E_{\rm hyd,HF}^{\rm QED,recoil}(1S)&=&
 \frac{8m_r^3\al^6}{3m_em_p}c_F^{(p)}\frac{m_e}{m_p}
 \left[
\kappa_p\left(\frac{7}{4} \ln \frac{1}{2 \alpha }-\ln 2+\frac{31}{36}\right)+2 \ln\frac{1}{2 \alpha }-6 \ln 2+\frac{65}{18}\right.\nn\\
&+&\left.\frac{\kappa_p }{\kappa_p+1}\left(-\frac{7}{4} \ln\frac{1}{2 \alpha }+4 \ln 2-\frac{31}{8}\right)
 \right]
 \,.
 \eea 
On top of these, there are recoil corrections of higher orders, as well as corrections due to the hadronic structure of the proton. It would be helpful to organize such computations/results using effective field theories techniques designed for few-body atomic physics such as NRQED \cite{Caswell:1985ui} and potential NRQED (see Refs.~\cite{Pineda:1997bj,Pineda:1997ie,Pineda:1998kn}). These theories profit from the hierarchy of scales that we have in the problem, which we name in the following way:
 \begin{itemize}
 \item 
 $m_e\al^2$: ultrasoft scale.
 \item
 $m_e\al$: soft scale.
 \item
 $m_e$, $m_r$: hard scale.
 \item
$m_{\pi}$, $\Delta=m_{\Delta}-m_p$, $m_{\mu}$: pion scale.
\item
$m_p$, $m_{\rho}$, $\Lambda_{\chi}$: chiral scale.
 \end{itemize}
 By considering ratios of these scales, the main expansion parameters are obtained:
\bea
&&
\label{ratio1}
\frac{m_{\pi}}{m_p} \sim \frac{m_{\mu}}{m_p} \approx \frac{1}{9}
\,,
\;
 \frac{m_r\al}{m_r}\sim \frac{m_r\al^2}{m_r\al}\sim \al \approx \frac{1}{137}
\,.
\eea

 The effects produced by the hard, pion and chiral scales are encoded in the Wilson coefficient $c_4^{p e}$ of the NRQED Lagrangian:
 \be
\delta {\cal L}_{Ne}^{\rm NR}=
-\displaystyle c_{4}^{p e}\frac{\al^2}{m_p^2} N_p^{\dagger}{\bfsigma}
  N_p \ {l}^{\dagger}_e{\bfsigma} l_e
\,.
\ee
Note that we have rescaled $c_4^{p e}$ by $\al^2$ compared with the definition used in Refs.~\cite{Pineda:2002as,Pineda:2002as2,Pineda:2004mx,Peset:2015zga}, to which we refer for the contextual discussion.

\begin{figure}[htb]
\begin{center}
\includegraphics[width=0.3\textwidth]{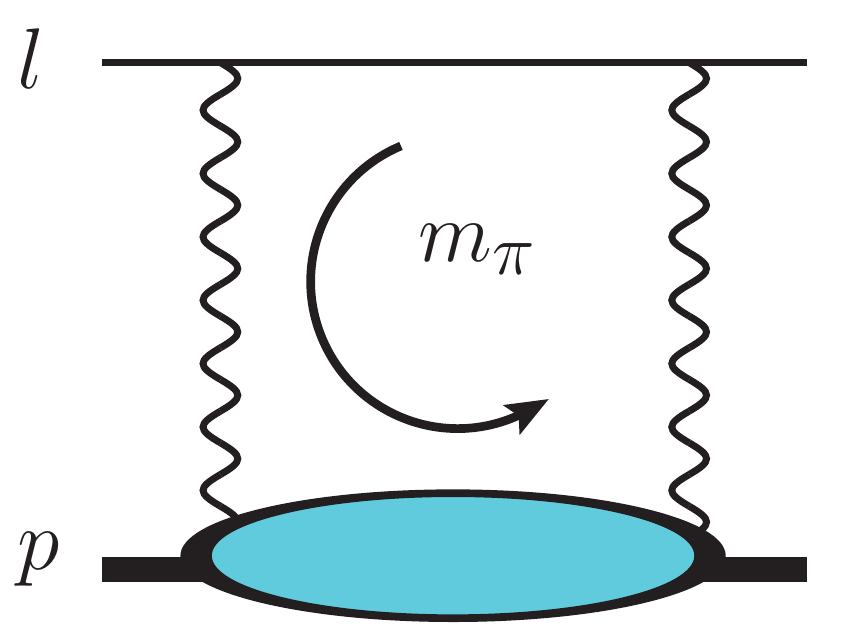}
\caption {\it Symbolic representation of $c_{4,\rm TPE}^{pl}$. The bubble 
represents the forward virtual-photon Compton 
tensor. It includes scales of the order of the lepton mass and higher (including the pion scale).}
\label{fig:c4}
\end{center}
\end{figure}

 The study of  $c_4^{p e}$ is one of the main subjects of this work. Like the magnetic moments, it has an expansion in powers of $\al$:
 \be
 c_4^{pe}\equiv c_{4,\rm TPE}^{pe}+\al\delta  c_4^{pe}
 \,,
 \ee
with $\delta  c_4^{pe} \sim 1+{\cal O}(\al)$, whereas $c_{4,\rm TPE}^{pe}$ is strict ${\cal O}(1)$ (although the number 
could be large), and has a one-to-one correspondence with the two-photon exchange energy shift of the hyperfine splitting. The coefficient $c_{4,\rm TPE}^{pe}$
can be written in a compact way (see Fig. \ref{fig:c4}) in 
terms of the structure functions of the forward virtual-photon Compton 
tensor. In Euclidean space it reads\footnote{Note that in the definition we give in this paper we take the structure functions as pure hadronic quantities, i.e. they do not include electromagnetic corrections.}
\bea
\nn
c_{4,\rm TPE}^{pe}&=&
\int {d^4k_E  \over 3\pi^2}
{1 \over k_E^2}{1 \over
k^4_E+4m_{e}^2k_{0,E}^2 }
\left\{
(k_{0,E}^2+2k_E^2)A_1(ik_{0,E},-k_E^2)+i3k_E^2{k_{0,E} \over m_p}A_2(ik_{0,E},-k_E^2)
\right\}
\,,
\\
\label{c4pe}
\eea
consistent with the expressions obtained long ago in Ref. \cite{DS}. 
$A_1$ and $A_2$ are the spin-dependent structure functions of the  forward virtual-photon Compton tensor,
\begin{equation}
 T^{\mu\nu} = i\!\int\! d^4x\, e^{iq\cdot x}
  \langle {p,s}| T \{J^\mu(x)J^\nu(0)\} |{p,s}\rangle
\,,
\end{equation}
which has the following structure ($\rho=q\cdot p/M_p\equiv v \cdot q$, although we will work in the 
rest frame where $\rho=q^0$):
\bea \label{inv-dec}
 T^{\mu\nu} &=
  &\left( -g^{\mu\nu} + \frac{q^\mu q^\nu}{q^2}\right) S_1(\rho,q^2) 
  + \frac1{m_p^2} \left( p^\mu - \frac{m_p\rho}{q^2} q^\mu \right)
    \left( p^\nu - \frac{m_p\rho}{q^2} q^\nu \right) S_2(\rho,q^2) 
	\\*
  && - \frac i{m_p}\, \epsilon^{\mu\nu\rho\sigma} q_\rho s_\sigma A_1(\rho,q^2)
	- \frac i{m_p^3}\, \epsilon^{\mu\nu\rho\sigma} q_\rho
   \bigl( (m_p\rho) s_\sigma - (q\cdot s) p_\sigma \bigr) A_2(\rho,q^2)
   \equiv T_S^{\mu\nu}+T_A^{\mu\nu}
   \nn
   \,.
\eea

Eq.~(\ref{c4pe}) keeps the complete dependence on $m_e$, the lepton mass, and admits a Taylor expansion in this parameter:
\be
c_{4,\rm TPE}^{pe}=c_{4,\rm TPE}^{pe,(0)}+\frac{m_e}{m_p}c_{4,\rm TPE}^{pe,(1)}+\cdots\,.
\ee
Note also that $c_{4,\rm TPE}^{pe,(0)}$ can be written in the following way:
\be
c_{4,\rm TPE}^{pe,(0)}=c_{4,\rm TPE}^{p,(0)}(\nu_{\rm pion})+\#\ln \frac{\nu_{\rm pion}}{m_e}
\,,
\ee
where $\nu_{\rm pion}$ is a factorization scale between $m_e$ and higher (pion and chiral) 
scales, and $c_{4,\rm TPE}^{p,(0)}(\nu_{\rm pion})$ does not depend on the lepton mass\footnote{$\nu_{\rm pion}$ is the ultraviolet cufoff of the theory named QED(e) in Refs.~\cite{Pineda:2002as,Pineda:2002as2}.}.

On the other hand the expansion of $\delta  c_4^{pe}$ starts with a negative power of $m_e$
\be
\delta  c_4^{pe}=\frac{m_p}{m_e}\delta  c_4^{pe,(-1)}+\delta  c_4^{pe,(0)}+\frac{m_e}{m_p}\delta  c_4^{pe,(1)}+\cdots
\,,
\ee
where~\cite{Kroll:1952zz}  
\be
\label{c4minus1}
\delta  c_4^{pe,(-1)}=\frac{2}{3}\pi Zc_F^{(p)}\left(\ln 2-\frac{5}{2}\right)
\ee 
follows from the 3rd term in Eq.~(\ref{QED}). Note that this expression includes effects associated to the electron vacuum polarization at the hard scale. $\delta  c_4^{pe,(0)}$ is logarithmically dependent on the ultraviolet cutoff scale of NRQED(e)/pNRQED, $\nu$. 
Such $\nu$-scale dependence cancels with the $\nu$-scale dependence of the computation in 
pNRQED yielding finite results (see Eq.~(\ref{recoil})). Similar considerations apply to subleading terms. 
Also, unlike $\delta  c_4^{pe,(-1)}$, $\delta  c_4^{pe,(0)}$ may receive 
contributions from the pion and chiral scales, i.e. from the hadronic structure of the proton. They are at present unknown and set the precision with which we can determine $c_{4,\rm TPE}^{pe}$, which is of ${\cal O}(\al)$ and so we estimate it to be of the order of 1\%. Overall, the structure of $\delta  c_4^{pe,(0)}$ is the following ($Z=1$)
\be
\delta  c_4^{pe,(0)}=
\frac{2}{3}\pi\left(2c_F^{(p)}+\frac{7}{4}\kappa_p^2\right)\ln \frac{m_e}{\nu}+K^{pe}
\,.
\ee
Note that $K^{pe}$ is scheme dependent. One may also consider splitting this term in the following way: 
$K^{pe}=K^{pe}_{\rm hard}+K^{pe}_{\rm had}$, 
where $K^{pe}_{\rm hard}$ encodes the effects associated to the hard scale exclusively. On the other 
hand $K^{pe}_{\rm had}$ encodes effects associated to the chiral and higher scales. Both coefficients will, in 
general, mix logarithmically: $K_{\rm hard} \sim \ln \nu_{\rm pion}/m_e$ and $K_{\rm had} \sim \ln \Lambda_{\chi}/\nu_{\rm pion}$ (for the discussion we do not distinguish between the pion and chiral scale), such that $K^{pe} \sim \ln \Lambda_{\chi}/m_e$. Therefore, all these coefficients are factorization-scale and scheme dependent. 

$c_4^{pe}$ is the Wilson coefficient that appears naturally in the computation of the hyperfine splitting:
\be
\delta E_{\rm hyd,HF}^{c_4}(nS)=\frac{4m_r^3Z^3\al^5}{\pi m_p^2n^3}c_{4}^{p e}
\,.
\ee
Therefore, this quantity can be determined with high precision by comparison with experiment 
after subtracting Eqs. (\ref{QED}) and (\ref{recoil}). We have to be careful though, as there are contributions associated to the hard scale that are incorporated in those equations. These comprise the effects associated to 
$\delta  c_4^{pe,(-1)}$ and to part of the hard part of $\delta  c_4^{pe,(0)}$. In other words, 
the following energy shift associated to $c_4^{pe}$ (in the following we take $Z=1$)
\be
\label{deltaEhard}
\delta E=\frac{4m_r^3\al^6}{\pi m_p^2n^3}
\left(
\frac{m_p}{m_e}\delta c_4^{pe,(-1)}+\frac{2}{3}\pi\left(2c_F^{(p)}+\frac{7}{4}\kappa_p^2\right)\ln \frac{m_e}{\nu}+\delta K_{\rm hard}
\right)
\ee
is included in Eqs.~(\ref{QED}) and (\ref{recoil}). $\delta K_{\rm hard}$ can be interpreted as a change of scheme in the renormalization of the Wilson coefficient. Note that this quantity is independent of the lepton up to corrections of ${\cal O}(m_e/m_p)$. 

\begin{figure}[htb]
\begin{center}
\includegraphics[width=0.45\textwidth]{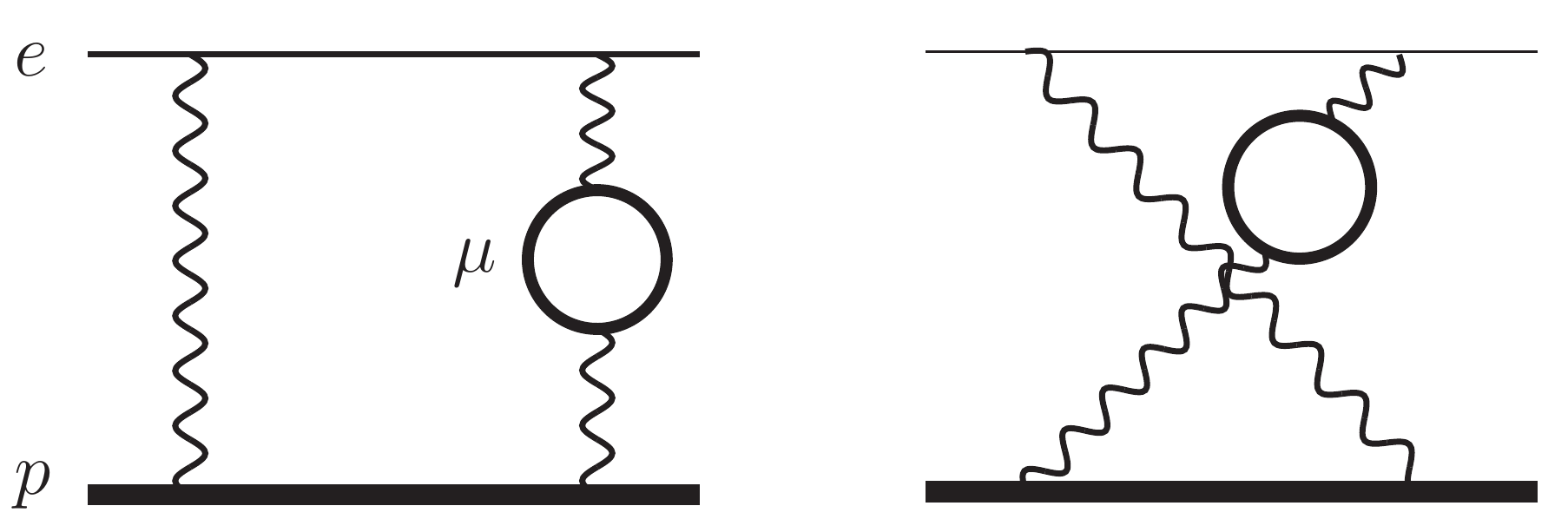}
\caption {\it Contribution to $K^{pe}$ due to muon vacuum polarization effects.}
\label{fig:Kpe}
\end{center}
\end{figure}

The difference between the experimental value of the hydrogen ground state 
hyperfine splitting and the pure QED computation then  reads
\bea
&&
\label{046}
E_{\rm hyd,HF}^{\rm exp}(1S)-\delta E_{\rm hyd,HF}^{\rm QED}(1S)-\delta E_{\rm hyd,HF}^{\rm QED, recoil}(1S)
\\
\nn
&=&\frac{4m_r^3\al^5}{\pi m_p^2}
\left(c_4^{pe}-\al\left(\frac{m_p}{m_e}\delta c_4^{pe,(-1)}+\frac{2}{3}\pi\left(2c_F^{(p)}+\frac{7}{4}\kappa_p^2\right)\ln \frac{m_e}{\nu}+\delta K_{\rm hard}\right)\right)+{\cal O}(m\al^8, m\al^7\frac{m_e}{m_p})
\\
\nn
&=&
-46.94(3)\;{\rm kHz},
\eea 
where the error comes from an estimate of the uncomputed ${\cal O}(\al^8, \al^7m_e/m_p)$ corrections to hyperfine splitting, as well as from the uncertainty of the measured value of the 
proton magnetic moment. 
From this result we determine (a specific part of) the Wilson coefficient 
of the four fermion operator of the NRQED Lagrangian (a preliminary number was already obtained in Refs.~\cite{Pineda:2002as,Pineda:2002as2} (see also \cite{Peset:2014jxa})), which we will name $\bar c_{4,\rm TPE}^{p e}$:\footnote{Note that effects associated to the diagram in Fig.~\ref{fig:Kpe}, due to the muon vacuum polarization, are included in $K^{pe}$, producing the correction $K^{pe}_{\mu,\rm VP}=\frac{\pi}{2}Zc_F^{(p)} \frac{m_p}{m_{\mu}}$~\cite{KarshemboimJPB}. This is natural as they correspond to corrections associated to the muon mass scale. In practice the generated energy shift is small: $\sim 0.27$ 
kHz.}
\bea
\label{c4peHF}
&&
\bar c_{4,\rm TPE}^{p e}\equiv c_4^{pe}-\al\left(\frac{m_p}{m_e}\delta c_4^{pe,(-1)}+\frac{2}{3}\pi\left(2c_F^{(p)}+\frac{7}{4}\kappa_p^2\right)\ln \frac{m_e}{\nu}+\delta K_{\rm hard}\right)
\\
\nn
&&=
c_{4,\rm TPE}^{p e}+\al [K^{pe}-\delta 
K_{\rm hard}]+{\cal O}(\al^2,\al\frac{m_e}{m_p})=-48.69(3)
\,.
\eea
It differs from $c_{4,\rm TPE}^{p e}$ by ${\cal O}(\al)$ effects. 
Note also that at ${\cal O}(\al)$ $c_4^{pe}$ is factorization-scale and scheme dependent: 
$c_4^{pe} \rightarrow c_{4,X}^{pe}(\nu)$. The factorization-scale dependence cancels with the logarithmic 
term explicitly displayed in Eq.~(\ref{c4peHF}). The scheme dependence of $c_{4,X}^{pe}(\nu)$ cancels with the scheme dependence of $\delta K^X_{\rm hard}$. Therefore, even if the number we obtain is definition dependent 
(we choose which part of the hard term is subtracted out), it can be taken as factorization-scale and scheme independent once the definition has been fixed. Related to that, we do not need the explicit expression 
of $\delta K_{\rm hard}$, which depends on the renormalization scheme used for the computation of the potential loops in 
the effective theory computation, to determine $\bar c_{4,\rm TPE}^{p e}$. Note also that the discussion is similar\footnote{For the proton radius the situation is somewhat worse, since one subtracts contributions at scales of the order of the proton mass.} for the definition of the proton radius (see~\cite{Pineda:2004mx}, and, for instance, Eq. (2.15) of \cite{,Peset:2015zga}): 
where one defines (at ${\cal O}(\al)$) the proton radius in terms of the factorization-scale and scheme dependent Wilson coefficient $c_D$ in the following way:
\be
r_p^2\equiv\frac{3}{4m_p^2}\left[c_{D,X}^{(p)}-1-\frac{4}{3}\frac{\al}{\pi}\ln\frac{m_p^2}{\nu^2}-\al\delta c_{D,X}\right]
\,,
\ee
 where $\delta c_{D,X}=0$ for $X=\MS$ scheme.
 
Finally, if we were specifically interested in $c_{4,\rm TPE}^{p e}$, the error is dominated by 
the unknown coefficient $\al K^{pe}$. Its natural size is $\al\times c^{pe}_4$. 
Assigning a $1\%$  error yields
\be
c_{4,\rm TPE}^{p e}=-48.7(5)
\,.
\ee

\section{Hyperfine splitting in muonic hydrogen}

We now discuss the computation of the hyperfine splitting in muonic hydrogen. A summary of results for the 
$2S$ hyperfine can already be found in \cite{Antognini:2013jkc} (see also~\cite{Indelicato:2012pfa}). Here the computation is organized as in Ref.~\cite{Peset:2015zga} (from which we also borrow the notation and refer for further details) but applied to the hyperfine splitting energy shift. Note also that the expressions below incorporate the muon magnetic moment, $c_F^{(\mu)}=1.00116592$,
 exactly wherever it appears (incorporating then higher order corrections in $\al$), since it corresponds to the Wilson coefficient that appears in the bilinear muon sector of the effective theory (as the binding energy of muonic hydrogen is much smaller than the muon and electron masses). 

The leading order correction reads (note that now $m_r=m_pm_{\mu}/(m_p+m_{\mu})$)
\be
\label{EFermi}
\delta E_{\rm Fermi}(nS)=
\frac{8m_r^3\alpha(Z\alpha)^3}{3m_pm_\mu n^3} c_F^{(p)}c_F^{(\mu)}
=182.6560379\frac{1}{n^3}\;{\rm meV}
\,.
\ee

Unlike in the hydrogen case there are corrections of ${\cal O}(m_r\al^5)$ due to the electron vacuum polarization. They show up as corrections from quantum mechanics perturbation theory and read
\begin{eqnarray}
\delta E_{\rm HF}^{V^{(2,1)}\times V_{\rm VP}^{(0,2)}}
&=&\begin{cases}\displaystyle 0.734488
\; {\rm meV} &{(1S)},\\0.074474
\; {\rm meV}&{(2S)}\,, 
\label{V21V0VP}\end{cases}
\end{eqnarray} 
\begin{eqnarray}
\delta E_{\rm HF}^{V_{\rm VP}^{(2,2)}}&=&\begin{cases}\displaystyle 0.374654\; {\rm meV} &{(1S)},\\ 0.048275 \; {\rm meV}&{(2S)}\,, 
\label{V22VP}\end{cases}
\end{eqnarray} 
where in both cases we pick up the part of $V^{(2,1)}$ and $V^{(2,2)}_{\rm VP}$ that contributes to the hyperfine (see Ref.~\cite{Peset:2015zga} for notation). Note that for $2S$ these numbers disagree with Pachucki 
\cite{pachucki1} but agree with Martynenko \cite{Martynenko:2004bt}.

Eqs. (\ref{EFermi}), (\ref{V21V0VP}) and (\ref{V22VP}) produce ${\cal O}(m_r\al^5)$ energy shifts due to scales below the muon mass. In these results we have kept the exact mass dependence.

Similarly to the hydrogen case, effects associated to the muon mass or higher scales are encoded in $c_4^{p\mu}$, the Wilson coefficient of the spin-dependent dimension six four-fermion operator. They 
produce the following correction
\be
\delta E_{\rm p\mu,HF}^{c_4}(nS)=\frac{4m_r^3Z^3\al^5}{\pi m_p^2n^3}c_{4}^{p\mu}
\,,
\ee
which starts contributing at ${\cal O}(\al^5)$. Following the discussion for hydrogen we write
\be
 c_4^{p\mu}\equiv c_{4,\rm TPE}^{p\mu}+\al\delta  c_4^{p\mu}
 \,,
 \ee
where $c_{4,\rm TPE}^{p\mu}$ is equal to Eq.~(\ref{c4pe}) after changing $m_e \rightarrow m_{\mu}$. 

$\delta  c_4^{p\mu}$ admits an expansion in powers of the muon mass as in the hydrogen case:
\be
\delta  c_4^{p\mu}=\frac{m_p}{m_\mu}\delta  c_4^{p\mu,(-1)}+\delta  c_4^{p\mu,(0)}+\frac{m_\mu}{m_p}\delta  c_4^{p\mu,(1)}+\cdots
\ee

The strict ${\cal O}(\al^5)$ contribution is associated to the two-photon exchange. It reads
\be
\delta E_{\rm p\mu,HF}^{\rm TPE}(nS)=\frac{4m_r^3Z^3\al^5}{\pi m_p^2n^3}c_{4,\rm TPE}^{p\mu}
\,.
\ee
This completes all the contributions to ${\cal O}(\al^5)$. Note that, formally, $\delta E_{\rm p\mu,HF}^{\rm TPE}$ has an extra $m_{\mu}/m_p$ 
suppression factor.

At ${\cal O}(\al^6)$ the dominant contributions are those obtained in the infinite proton mass limit. For the soft and ultrasoft scale computations, this limit makes the system equivalent to hydrogen but with an active light fermion. 
The Breit corrections are equal to the hydrogen case. They read (see \cite{Eides:2000xc} for instance)
\begin{eqnarray}
\delta E_{Breit}&=&\frac{8m_r^3\alpha(Z\alpha)^3}{3m_pm_\mu} c_F^{(p)}
Z^2\alpha ^2 
\begin{cases}\displaystyle \frac{3}{2}\; {\rm meV} &{(1S)},\\\frac{17}{64}\; {\rm meV}&{(2S)}\,.
\label{Breit}\end{cases}
\end{eqnarray} 

For the hard scale (which now we take to be of ${\cal O}(m_{\mu})$) there is an important change. The hard and pion scales are approximately equal, and, a priori, they should be computed at once. This means that the separation we made for the case of hydrogen between the hard and the pion scales has to be reanalyzed. Chiral counting suggests that chiral corrections always have a $1/\Lambda_{\chi}^2$ suppression factor. Therefore, there is no contribution to $\delta c_4^{p\mu,(-1)}$ at any order in $\al$ associated to those scales, which is then completely determined by the point-like computation. Neglecting 
electron vacuum polarization effects, one has the 
same expression as for the hydrogen case (Eq. (\ref{c4minus1})):  
\be
\delta  c_4^{p\mu,(-1)}=\frac{2}{3}\pi Zc_F^{(p)} \left(\ln 2-\frac{5}{2}\right)
\,,
\ee 
which produces the following correction to the energy shift:
\be
\label{dv}
\delta E^{V^{(2,3)}}\Bigg|_{\delta c_4^{(-1)}}=
\frac{8m_r^3\alpha(Z\alpha)^3}{3 m_pm_\mu n^3} c_F^{(p)}\alpha ^2 Z \left[\ln 2-\frac{5}{2}\right]
\,.
\ee

Eqs.~(\ref{Breit}) and (\ref{dv}) are all the ${\cal O}(m_{\mu}\al^6)$ corrections for a hydrogen-like system in the infinite proton mass limit. Nevertheless, electron vacuum polarization effects also produce corrections to the energy at this order.  At present the complete set of these corrections is unknown. Therefore, one cannot claim that the complete ${\cal O}(m_{\mu}\al^6)$ correction to the hyperfine splitting of muonic hydrogen is known. Nevertheless, vacuum polarization effects seem to be suppressed compared with standard relativistic corrections for the same power of $\alpha$. To test this idea we have performed a partial analysis by evaluating some higher order corrections generated by the electron vacuum polarization. We first take those corrections generated by second-order quantum mechanics perturbation theory of the spin-dependent Breit-Fermi potential with the two-loop correction to the static potential. The result reads
\begin{eqnarray}
\delta E_{\rm HF}^{V^{(2,1)}\times V_{\rm VP}^{(0,3)}}
&=&\begin{cases}\displaystyle 
0.00556\; {\rm meV} &{(1S)},
\\
0.00056\; {\rm meV}&{(2S)}\,. 
\label{V21V0VP3}\end{cases}
\end{eqnarray} 
The other correction we compute is the one associated to the two-loop correction to the spin-dependent Breit-Fermi potential produced by the electron vacuum polarization effects. It reads
\begin{eqnarray}
\delta E_{\rm HF}^{V_{\rm VP}^{(2,3)}}&=&\begin{cases}\displaystyle 
2.9168\times 10^{-3}
\; {\rm meV} &{(1S)},\\ 
3.7455\times 10^{-5}
\; {\rm meV}&{(2S)}\,. 
\label{V23VP}\end{cases}
\end{eqnarray} 
We indeed observe that they are suppressed. For the $2S$ our numbers agree with Borie~\cite{Borie:2012zz}. 
We add both results to our final numbers and use their size for the estimate of the error associated to the yet uncalculated other 
${\cal O}(m_{\mu}\al^6)$ electron vacuum polarization corrections.

 \begin{figure}[htb]
\begin{center}
\includegraphics[width=0.45\textwidth]{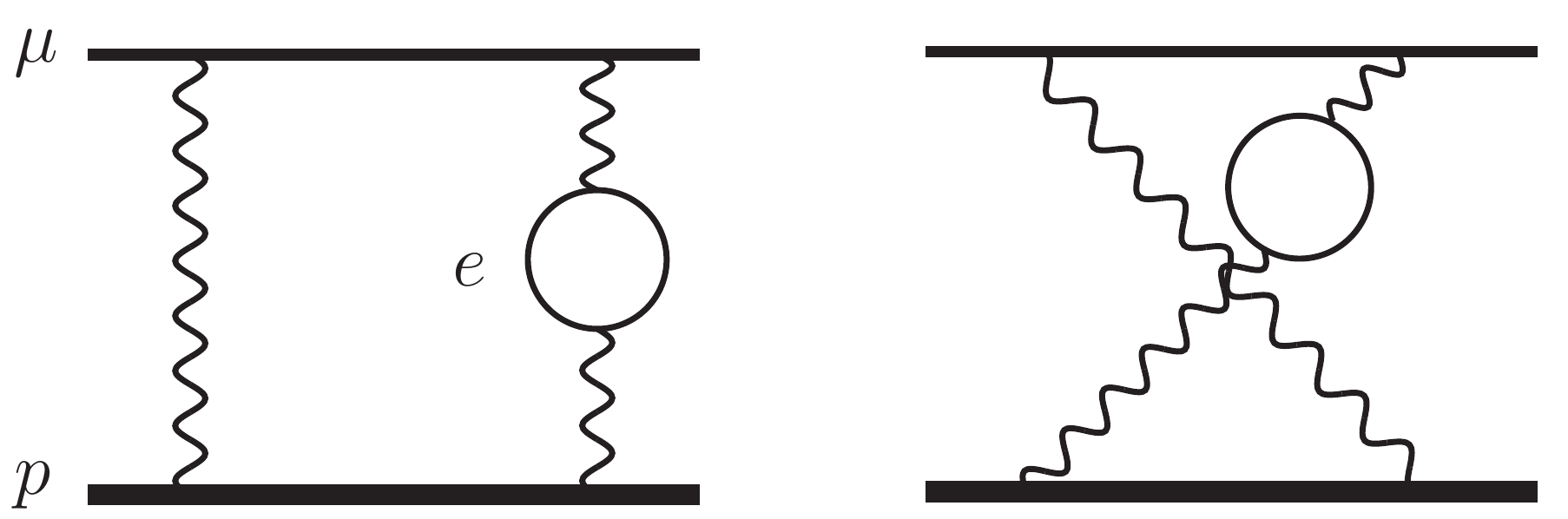}
\caption {\it Contribution to $\delta c_4^{p\mu}$ due to electron vacuum polarization effects.}
\label{fig:deltac4pmu}
\end{center}
\end{figure}

A priori, electron vacuum polarization corrections can also contribute to $\delta c_4^{p\mu,(-1)}$, due to the diagram in Fig.~\ref{fig:deltac4pmu} (note that we only consider energy scales of the order of the mass of the muon or higher, lower scales are subtracted, as they are incorporated in the effective theory computation. See also the discussion in Ref.~\cite{Martynenko:2004bt}). We have performed a numerical computation of that diagram in the point-like approximation, which does not appear to contribute to $\delta c_4^{p\mu,(-1)}$. We have found a very tiny energy shift $\delta E(2S) \sim 4\times 10^{-5}$ meV, which should be interpreted as a recoil correction. This connects us with the other set of corrections that we consider: The recoil corrections of ${\cal O}(m\al^6\frac{m_{\mu}}{m_p})$.  For the ground state the leading correction corresponds to Eq.~(\ref{recoil}) changing the electron by the muon mass. This is the expression we use in Table \ref{table1S} for entry ix). 
For the $2S$ state one can deduce the analogous expression from the $8E_{\rm hyd,HF}(2S)-E_{\rm hyd,HF}(1S)$ energy shift theoretical expression (see \cite{Karshenboim:2005iy}) and Eq.~(\ref{recoil}). One obtains (for $Z=1$)
\bea
 \label{recoillog}
 \delta E_{\rm HF}^{\rm QED,recoil}(2S)&=&
 \frac{m_r^3\al^6}{3m_{\mu}m_p}\frac{m_{\mu}}{m_p}
 \left[\frac{1559}{1152}-\frac{1735}{576} \kappa_p-\frac{313}{1152}\kappa_p^2-\left(\frac{15 \kappa_p^2}{8}+\frac{11 \kappa_p}{4}+\frac{57}{8}\right) \ln2\right.\nn\\
 &+&\left.\left(2c_F^{(p)}+\frac{7}{4}\kappa_p^2
 \right)\ln \frac{1}{ \alpha }\right]
 \,.
 \eea 
Note that, for instance, in Ref.~\cite{Carlson:2008ke} this contribution is obtained rescaling the $1S$ result by $1/2^3$. This is not correct, as it is a bound state computation, and the dependence on the principal quantum number, $n$, is more complicated. On the other hand the logarithmic correction, the ${\cal O}(m\al^6\frac{m_{\mu}}{m_p}\ln\al)$ term, is the same, i.e. it only has the $1/2^3$ factor difference with respect to
 the $1S$ result. The difference in the finite piece introduces an error in their computation, which is, nevertheless, small. As this difference is associated to the point-like part of the computation, we may expect that its size is small compared to pure hadronic effects for the same power of $\al$.

Even if Eq.~(\ref{recoillog}) has a $\frac{m_{\mu}}{m_p}$ suppression factor, it is also logarithmically enhanced, making this contribution roughly equivalent to the contribution due to Eqs.~(\ref{Breit}) and (\ref{dv}) (see Tables \ref{table1S} and \ref{table2S}). 

Finally, we compute the ${\cal O}(\al)$ correction to the energy shift proportional to $c_4^{p\mu}$ produced by 2nd order perturbation theory with $V_{\rm VP}^{(0,2)}$, the electron vacuum polarization correction to the static potential (for the precision of the computation we can replace $c_4^{p\mu}$ by $c_{4,\rm TPE}^{p\mu}$):
\begin{eqnarray}
\delta E_{\rm HF}^{V^{(2,2)}\times V_{\rm VP}^{(0,2)}}\Bigg|_{c_4^{p\mu}}
&=&
\frac{4m_r^3Z^3\al^5}{\pi m_p^2}c_4^{p\mu}
\begin{cases}\displaystyle \quad 
4.02115\times10^{-3}&{(1S)},\\\; \frac{1}{8}\times
3.26181\times10^{-3}
&{(2S)}\,.
\label{V22c4V0VP}\end{cases}
\end{eqnarray} 
This produces a tiny correction. 

As in the hydrogen case (see Eq.~(\ref{deltaEhard})), there is some double counting in the above computations,  and (in the following we take $Z=1$)
\be
\delta E=\frac{4m_r^3\al^6}{\pi m_p^2n^3}
\left(
\frac{m_p}{m_{\mu}}\delta c_4^{p\mu,(-1)}+\frac{2}{3}\pi\left(2c_F^{(p)}+\frac{7}{4}\kappa_p^2\right)\ln \frac{m_{\mu}}{\nu}+\delta K_{\rm hard}
\right)
\ee 
is already included in Eqs.~(\ref{dv}) and (\ref{recoillog}), where we have used 
\be
\delta  c_4^{p\mu,(0)}=
\frac{2}{3}\pi\left(2c_F^{(p)}+\frac{7}{4}\kappa_p^2\right)\ln \frac{m_\mu}{\nu}+K^{p\mu}
\,.
\ee 
Nevertheless, what is important is that the same $\delta K_{\rm hard}$ applies to the hydrogen and muonic hydrogen and for different quantum numbers, which allows us to relate different measurements. 
 
We collect all the results of this section in Tables \ref{table1S} and \ref{table2S}, define\footnote{$K^{p\mu}$ includes all physical effects associated to the chiral and higher scales, except for those we explicitly subtract from point-like computations at the muon mass scale. This is different for $K^{pe}$, which incorporates all the corrections generated by the chiral scale.}
\bea
\bar c_{4,\rm TPE}^{p \mu}&\equiv&
c_{4}^{p \mu}-\al\left(
\frac{m_p}{m_{\mu}}\delta c_4^{p\mu,(-1)}+\frac{2}{3}\pi\left(2c_F^{(p)}+\frac{7}{4}\kappa_p^2\right)\ln \frac{m_{\mu}}{\nu}+\delta K_{\rm hard}
\right)
\\
\nn
&=&
c_{4,\rm TPE}^{p \mu}+\al [K^{p\mu}-\delta 
K_{\rm hard}]+{\cal O}(\al^2,\al\frac{m_{\mu}}{m_p})
\,,
\\
\delta \bar E^{\rm TPE}_{p \mu,\rm HF}(nS)&=&\frac{4m_r^3\al^5}{n^3 \pi m_p^2}\bar c_{4,\rm TPE}^{p \mu}
\,,
\eea
and obtain for the $1S$ hyperfine splitting
\be
\label{1Smuonichydrogen}
E_{p\mu,\rm HF}^{\rm th}(1S)=183.788(7)+(1+0.0040)\delta \bar E^{\rm TPE}_{p \mu,\rm HF}(1S)\,,
\ee
where the error in the first term is the expected size of the ${\cal O}(\al^6)$ uncomputed corrections (either 
related to the electron vacuum polarization effects or to higher recoil corrections along the lines of item ix) in Table \ref{table1S}).

For the $2S$ we obtain
\bea
\label{2Smuonichydrogen}
&&E_{p\mu,\rm HF}^{\rm th}(2S)=22.9579(8)+(1+0.0033)\delta \bar E^{\rm TPE}_{p \mu,\rm HF}(2S)\,,
\eea
where again the error in the first term is the expected size of the ${\cal O}(\al^6)$ uncomputed corrections (either 
related to the electron vacuum polarization effects or to higher recoil corrections along the lines of item ix) in Table \ref{table2S}).
 
\begin{table}[ht]
\addtolength{\arraycolsep}{0.15cm}
$$
\begin{array}{|lc|c|c|c   |r l|}
 \hline 
&{\rm i)}&{\cal O} (m_r \alpha^4)& \delta E_{\rm Fermi} &  {\rm Eq.}\; (\ref{EFermi}) & 182.&\hspace{-0.5cm}65604  
\\ \hline
&{\rm ii)}&{\cal O} (m_r \alpha^5)& V^{(2,1)}\times  V^{(0,2)}_{\rm VP} & {\rm Eq.}\; (\ref{V21V0VP}) &0.&\hspace{-0.5cm}73449 
\\ \hline
&{\rm iii)}&{\cal O} (m_r \alpha^5)& V^{(2,2)}_{\rm VP} & {\rm Eq.}\; (\ref{V22VP}) &0.&\hspace{-0.5cm}37465
\\ \hline
&{\rm iv)}&{\cal O} (m_r \alpha^5\frac{m_{\mu}}{m_p})& \delta \bar E^{\rm TPE} & {\rm Eq.}\; (\ref{TPEpepmu})& -1.&\hspace{-0.5cm}161(20)
\\ \hline
&{\rm v)}&{\cal O} (m_r \alpha^6)& V^{(2,1)}\times  V^{(2,1)}  & {\rm Eq.}\; (\ref{Breit})&   0.&\hspace{-0.5cm}01457
\\ \hline
&{\rm vi)}&{\cal O} (m_r \alpha^6)& V^{(2,3)}; \delta c_4^{(-1)} & {\rm Eq.}\; (\ref{dv})&  -0.&\hspace{-0.5cm}01755
\\ \hline
&{\rm vii)}&{\cal O} (m_r \alpha^6)& 
V^{(2,1)}\times  V^{(0,3)}_{\rm VP}  & {\rm Eq.}\; (\ref{V21V0VP3}) &   0.&\hspace{-0.5cm}00556
\\ \hline
&{\rm viii)}&{\cal O} (m_r \alpha^6)& V^{(2,3)}_{\rm VP} & {\rm Eq.}\; (\ref{V23VP}) &   0.&\hspace{-0.5cm}00292
\\ \hline
&{\rm ix)}&{\cal O} (m_r \alpha^6
\times \frac{m_{\mu}}{m_{p}})& V^{(2,1)}\times  V^{(2,1)}  &  {\rm Eq.}\; (\ref{recoil})& 0.&\hspace{-0.5cm}01752
\\ \hline
&{\rm x)}&{\cal O} (m_r \alpha^6
\times \frac{m_{\mu}}{m_{p}})& V^{(2,2)}\times  V^{(0,2)}_{\rm VP};c_{4,\rm TPE} & {\rm Eq.}\; (\ref{V22c4V0VP}) &-0.&
\hspace{-0.5cm}  
00466
\\[0.05cm] \hline \hline 
& &{\rm Total\; sum}&  & {\rm Eq.}\; (\ref{HFfinal}) & 182.&\hspace{-0.5cm}623(27)
\\ \hline
\end{array}
$$
\caption{{\it The different contributions to the hyperfine splitting for the 1S in muonic hydrogen in } meV {\it units. The value for the iv) and x) entries is obtained from the hydrogen hyperfine measurement.}}
\label{table1S}
\end{table}

\begin{table}[htb]
\addtolength{\arraycolsep}{0.15cm}
$$
\begin{array}{|lc|c|c|c   |r l|}
 \hline 
&{\rm i)}&{\cal O} (m_r \alpha^4)& \delta E_{\rm Fermi} &  {\rm Eq.}\; (\ref{EFermi}) & 22.&\hspace{-0.5cm}832005  
\\ \hline
&{\rm ii)}&{\cal O} (m_r \alpha^5)& V^{(2,1)}\times  V^{(0,2)}_{\rm VP} & {\rm Eq.}\; (\ref{V21V0VP}) &0.&\hspace{-0.5cm}074474  
\\ \hline
&{\rm iii)}&{\cal O} (m_r \alpha^5)& V^{(2,2)}_{\rm VP} & {\rm Eq.}\; (\ref{V22VP}) &0.&\hspace{-0.5cm}048275
\\ \hline
&{\rm iv)}&{\cal O} (m_r \alpha^5\frac{m_{\mu}}{m_p})& \delta \bar E^{\rm TPE} & {\rm Eq.}\; (\ref{TPEpepmu})& -0.&\hspace{-0.5cm}1451(25)
\\ \hline
&{\rm v)}&{\cal O} (m_r \alpha^6)& V^{(2,1)}\times  V^{(2,1)} & {\rm Eq.}\; (\ref{Breit})&   0.&\hspace{-0.5cm}002581
\\ \hline
&{\rm vi)}&{\cal O} (m_r \alpha^6)& V^{(2,3)}; \delta c_4^{(-1)} & {\rm Eq.}\; (\ref{dv})&  -0.&\hspace{-0.5cm}002194
\\ \hline
&{\rm vii)}&{\cal O} (m_r \alpha^6)& 
V^{(2,1)}\times  V^{(0,3)}_{\rm VP}  & {\rm Eq.}\; (\ref{V21V0VP3}) &   0.&\hspace{-0.5cm}000375
\\ \hline
&{\rm viii)}&{\cal O} (m_r \alpha^6)& V^{(2,3)}_{\rm VP} & {\rm Eq.}\; (\ref{V23VP}) &   0.&\hspace{-0.5cm}000563
\\ \hline
&{\rm ix)}&{\cal O} (m_r \alpha^6
\times \frac{m_{\mu}}{m_{p}})& V^{(2,1)}\times  V^{(2,1)}  &  {\rm Eq.}\; (\ref{recoillog})& 0.&\hspace{-0.5cm}001846
\\ \hline
&{\rm x)}&{\cal O} (m_r \alpha^6
\times \frac{m_{\mu}}{m_{p}})& V^{(2,2)}\times  V^{(0,2)}_{\rm VP};c_{4,\rm TPE} & {\rm Eq.}\; (\ref{V22c4V0VP}) &-0.&
\hspace{-0.5cm}  
000473
\\[0.05cm] \hline \hline 
& &{\rm Total\;sum}&  & {\rm Eq.}\; (\ref{HFfinal}) & 22.&\hspace{-0.5cm}8123(33)
\\ \hline
\end{array}
$$
\caption{{\it The different contributions to the hyperfine splitting for the 2S in muonic hydrogen in } meV {\it units. The value for the iv) and x) entries is obtained from the hydrogen hyperfine measurement.}}
\label{table2S}
\end{table}

Finally, we give a prediction for 
\be
8E_{p\mu,\rm HF}^{\rm th}(2S)-E_{p\mu,\rm HF}^{\rm th}(1S)=-0.124(3)\;{\rm meV}
\,.
\ee
This number is in good agreement with the number quoted in Ref.~\cite{Martynenko:2004bt}. We have checked, for the 1S contributions related to the vacuum polarization, that we agree with the analytic expressions obtained in Ref.~\cite{Martynenko:2004yg} but the numerical agreement is not very good.

\subsection{Determination of $c_{4,\rm TPE}^{p\mu}$ from hyperfine splitting in muonic hydrogen}

The possibility of measuring the hyperfine splitting in muonic hydrogen has become a reality in recent years: \cite{Pohl:2010zza,Antognini:1900ns,Sato:2015gra,Adamczak:2016pdb}. Actually an experimental number has been given for the $2S$ hyperfine splitting in Ref.~\cite{Antognini:1900ns}\footnote{After using a theoretical value for the fine splitting.}. 
In addition, there is an ongoing effort at PSI to determine the hyperfine splitting of the ground state at around the 100 MHz level accuracy~\cite{Aldo}. This calls for a determination of $c_4^{p\mu}$ as precise as possible. One possibility is to fix it by direct comparison to the experimental results from muonic hydrogen physics: 
\be
E_{\mu p,\rm HF}^{\rm exp}(2S)=22.8089(51) \; {\rm meV}  
\,.
\ee

Using this number and Eq.~(\ref{2Smuonichydrogen}) we obtain
\be
\label{2Smuon}
\delta \bar E^{\rm TPE}_{p \mu,\rm HF}(2S)
=-0.1490\,(51)_{\rm exp}(8)_{\rm th}
\,,
\ee
for the two-photon exchange contribution and
\be
\bar c_{4,\rm TPE}^{p\mu}=- 46.5\,(1.6)_{\rm exp}(0.3)_{\rm th}
\,,
\ee
for the Wilson coefficient. The errors are dominated by experiment.  
On the other hand, the foreseen determination of the hyperfine splitting of the ground state of muonic hydrogen would eliminate most of this error. 

\subsection{Determination of $\bar c_{4,\rm TPE}^{p\mu}$ from hyperfine splitting in hydrogen}

There is an alternative way to determine $c_4^{p\mu}$. One can actually predict this number from theory with the help of the hyperfine hydrogen measurement. The coefficients $c_4^{p\mu}$ and $c_4^{pe}$ can be related to a large extent (see the discussion in 
Refs.~\cite{Pineda:2002as,Pineda:2002as2}). Obviously, one can always write
\bea
\label{evsmuon}
\bar c_{4,\rm TPE}^{p\mu}
&=&\bar c_{4,\rm TPE}^{pe}+(\bar c_{4,\rm TPE}^{p\mu}-\bar c_{4,\rm TPE}^{pe})
\\
\nn
&=&\bar c_{4,\rm TPE}^{pe}+\left([c_{4,\rm TPE}^{p{\mu}}-c_{4,\rm TPE}^{p{e}}]+\al [K^{p\mu} -K^{pe}]\right)
+{\cal O}(\al\frac{m_{\mu}}{m_p})
\,.
\eea
The first term can be determined from the hyperfine energy shift measurement with high precision. The constants $K$ that appear in the above equation are unknown at present. As for the hydrogen case, they introduce an error of around 1\%, which we will add to the error budget. We then approximate 
Eq. (\ref{evsmuon}) to 
\be
c_{4,\rm TPE}^{p \mu}=c_{4,\rm TPE}^{p e}
+[c_{4,\rm TPE}^{p{\mu}}-c_{4,\rm TPE}^{p{e}}]
+{\cal O}(\al)
\,.
\ee
The key point is that the second term within parenthesis can be determined using chiral perturbation theory. 

As we have already mentioned, $c_{4,\rm TPE}^{p e}$ and 
$c_{4,\rm TPE}^{p \mu}$ encode the  contribution associated to the diagram shown in Fig.~\ref{fig:c4} shifting the external lepton ($l=e,\mu$). In this diagram all scales from the mass of the lepton, $m_l$, up to infinity are incorporated, whereas smaller scales are cut off, as those effects are associated to the bound state 
dynamics which is taken into account in the computation in the pNRQED effective theory. These coefficients keep the complete dependence 
on $m_{l}$ and are valid both for NRQED($\mu$) and NRQED($e$), 
i.e. for hydrogen and muonic hydrogen. 
Their computation can be organized in the following way
\be
\label{c4split}
c_{4,\rm TPE}^{pl}=
c_{4,\rm R}^{pl}
+ c_{4,\rm point-like}^{pl}
+ c_{4,\rm Born}^{pl}+ c_{4,\rm pol}^{pl} 
\,.
\ee
Within the effective field theory framework the contribution from energies of ${\cal O}(m_\rho)$
or higher in Eq. (\ref{c4split}) are encoded in $c_{4,\rm R}^{pl}\simeq
c_{4,\rm R}^{p}$. The other terms (associated to energies of ${\cal O}(m_\pi)$) can be computed using chiral effective theories for heavy baryons~\cite{Jenkins:1990jv}. Indeed they were computed with logarithmic 
accuracy: ${\cal O}(\ln m_{\pi}^2, \ln (M_{\Delta}-M_p), \ln m_{l}))$ in Refs.~\cite{Pineda:2002as,Pineda:2002as2}.  Even 
though they diverge logarithmically such divergence cancels in the difference $c_{4,\rm TPE}^{p{\mu}}-c_{4,\rm TPE}^{p{e}}$. 

A chiral computation would allow us to relate $c_4^{p{\mu}}$ and $c_4^{p{e}}$ in a model independent way. Since $c_{4,\rm R}^{pl}\simeq c_{4,\rm R}^{p}$ up to terms of ${\cal O}( m_{l}/\lQ)$, 
we can obtain the following relation 
\be
\label{deltac4mue}
c_{4,\rm TPE}^{p{\mu}}-c_{4,\rm TPE}^{p{e}}=\left[c_{4,\rm Born}^{p{\mu}}-c_{4,\rm Born}^{p{e}}\right]+
\left[c_{4,\rm point-like}^{p{\mu}}-c_{4,\rm point-like}^{p{e}}\right]+ \left[c_{4,\rm pol}^{p{\mu}}
- c_{4,\rm pol}^{p{e}}\right] 
\,.
\ee
and compute each of these terms. 
Actually, such computation was already done in Ref.~\cite{Peset:2014jxa} at leading order in the chiral expansion. Here we further elaborate on that result by considering the leading corrections to that computation and adding an error analysis to the final results\footnote{For the numerical values of the coefficients we use those of Ref.~\cite{Peset:2014jxa} except for $g_A=1.2723(23)$, which we take from Ref.~\cite{Olive:2016xmw}, $b_{1,F}=3\kappa_p/\sqrt{2}$ and $g_{\pi N\Delta}=3/(2\sqrt{2})g_A$, which we have fixed to the large $N_c$ prediction.}. 

For the point-like contribution we obtain
\bea
\nn
c_{4,\rm point-like}^{p\mu}-c_{4,\rm point-like}^{p{e}}
&=&
\left(1-\frac{ \kappa^2_p}{4}\right)
\ln{m_{\mu}^2\over m_e^2}
+\frac{m_{\mu}^2}{m_p^2} \left(1 + \frac{\kappa_p}{2} (1 - \frac{\kappa_p}{6}) \right) \ln\frac{m_{\mu}^2}{\nu_{\rm pion}^2}     
\\
\label{pointlike}
&\simeq& 2.09 -0.09=2.00(9)
\,,
\eea
where we use the $\frac{m_{\mu}^2}{m_p^2}$ logarithmic term for the estimate of subleading terms setting the factorization scale to the proton mass. We observe a nice convergence pattern.

The polarizability term was computed using the expression of the spin-dependent structure functions obtained in \cite{Ji:1999mr,Peset:2014jxa}. We obtain (note that this term vanishes in the large $N_c$ limit, except for the tree-level-like contribution)
\begin{eqnarray}
c_{4,\rm pol}^{p{\mu}}- c_{4,\rm pol}^{p{e}}
&=&\begin{cases}\displaystyle 0.17(9) &(\pi),\\ 0.25(10)&(\pi\&\Delta)\,, 
\label{pol}
\end{cases}
\end{eqnarray} 
where we use the same error analysis as in Ref. \cite{Peset:2014jxa}. It is possible to compare this number with a determination using dispersion relations~\cite{Carlson:2008ke}. They compute both coefficients, for hydrogen and muonic hydrogen. For the difference they obtain $c_{4,\rm pol}^{p{\mu}}- c_{4,\rm pol}^{p{e}}=-0.3(1.4)$, where we have combined in quadrature the error of the individual coefficients. Note that there is a very strong cancellation between both terms, so small inaccuracies in the parameterization of the experimental data could get amplified, still, within errors, there is perfect agreement. We should also mention that a recent analysis using relativistic baryon chiral perturbation theory has challenged standard values for the polarizability term obtained from dispersion relations \cite{Hagelstein}. 

For the Born term we get zero at leading order (in other words, the Zemach correction \cite{Zemach:1956zz} is the same for the hydrogen and muonic hydrogen). 
The first nontrivial term in the $1/m_p$, $1/F_{\pi}$ expansion reads (see \cite{pachucki1} for instance)
\bea
\label{c4Born}
&&
c_{4,\rm Born}^{p\mu}-c_{4,\rm Born}^{pe}
=
-\int_0^{\infty}dp\frac{1}{3 p} G^{(1)}_M(-p^2)
\\
&&
\times
\left[
\left(\frac{p^2
   \kappa_p}{m_{\mu}^2}+\frac{32 m_\mu^4-8 m_\mu^2 p^2
   (\kappa_p+2)-2 p^4 \kappa_p}{m_\mu^2 p
   \left(\sqrt{4 m_\mu^2+p^2}+p\right)}+8\right)-
   \left(m_{\mu}\rightarrow m_e\right)
   \right]
   \,,
   \nn
\eea
where we have approximated the magnetic Sachs form factor to 
$G^{(1)}_M(-p^2)$, its one-loop chiral expression~\cite{Gasser:1987rb,Bernard:1992qa,Bernard:1998gv}. 
The integral in Eq.~(\ref{c4Born}) is finite. The high energy behavior is cut by the cancellation between the integrands of the muon and electron in such a way that the ultraviolet behavior of the integrand scales as $1/p^2$ (note that $G^{(1)}_M(-p^2) \sim p$ for large $p$)\footnote{This only happens for the difference, the individual terms are actually power-like divergent.
}.
 The integrand in the $p \rightarrow 0$ is also finite since $G^{(1)}_M(-p^2) \sim p^2$ for small $p$. After numerical evaluation we find (we use the same error analysis as in Ref. \cite{Peset:2014jxa})
\begin{eqnarray}
\label{Bornchiral}
c_{4,\rm Born}^{p{\mu}}- c_{4,\rm Born}^{p{e}}
&=&\begin{cases}\displaystyle 0+1.11(55) &(\pi),\\ 0+1.42(53)&(\pi\&\Delta)\,.
\end{cases}
\end{eqnarray}  
Even if formally subleading this contribution turns out to be more important than the polarizability contribution. There are reasons for that, based on large $N_c$ arguments, already discussed in Refs.~\cite{Pineda:2002as,Pineda:2002as2}. On the other hand, Eq.~(\ref{Bornchiral}) is much smaller than the typical size of $c_4 \sim -45$, showing a nice convergence pattern. The sum of Eq.~(\ref{pointlike}) and (\ref{Bornchiral}): 3.42(62), can be compared with 
a determination using dispersion relations~\cite{Carlson:2008ke}, similarly as the polarizability case. They obtain 4.07(21), where we combine the errors of the individual coefficients in quadrature. Within one sigma it agrees with our prediction.

Overall, combining the three contributions, Eqs.~(\ref{pointlike}), (\ref{pol}) and (\ref{Bornchiral}), we obtain 
\be
\label{TPEdifference}
c_{4,\rm TPE}^{p{\mu}}-c_{4,\rm TPE}^{p{e}}=3.68(72)
\,.
\ee
Again this number can be compared with the determination from dispersion relations~\cite{Carlson:2008ke}: 3.8(1.4). One gets perfect agreement.

Using Eqs.~(\ref{c4peHF}), (\ref{TPEdifference}) and Eq.~(\ref{evsmuon}) we obtain
 \begin{eqnarray}
\label{TPEpepmu}
\bar c_{4,\rm TPE}^{p{\mu}}= -45.30(77) 
\;,
\qquad \delta \bar E^{\rm TPE}_{p \mu,\rm HF}
&=&\begin{cases}\displaystyle -1.161(20) &(1S),\\ -0.1451(25) &(2S)\,, 
\end{cases}
\end{eqnarray}  
where we take $\al(K^{p\mu} -K^{pe})=-\al K^{pe}_{\mu,\rm VP}$ as the estimate of this term for the numerical evaluation. 
The error is the combination in quadrature of the error in Eq.~(\ref{TPEdifference}), the error associated to 
$\al(K^{p\mu} -K^{pe})$ (which we take of the order of $\al \times (c_{4,\rm TPE}^{p{\mu}}-c_{4,\rm TPE}^{p{e}})$ and $K^{pe}_{\mu,\rm VP}$), and the error of Eq.~(\ref{c4peHF}). The last two errors are clearly subdominant compared to the error in Eq.~(\ref{TPEdifference}). Note that Eq.~(\ref{TPEpepmu}) has no input from muonic hydrogen data. 

Similarly to the two-photon exchange contribution $c_{4,\rm TPE}$, the constants $K$ also admit a computation in the chiral theory such that $K^{p\mu} -K^{pe}=-K^{pe}_{\mu,\rm VP}+
K^{p\mu}_{\chi PT} -K^{pe}_{\chi PT}+{\cal O}(\frac{m_{\mu}}{m_p})$. In other words, the leading term in the chiral expansion can be 
completely determined from a chiral  plus point-like computation without any new counterterm. Therefore, there is room to improve this analysis, were this necessary.

\begin{figure}[h]
\begin{center}
\includegraphics[width=0.82\textwidth]{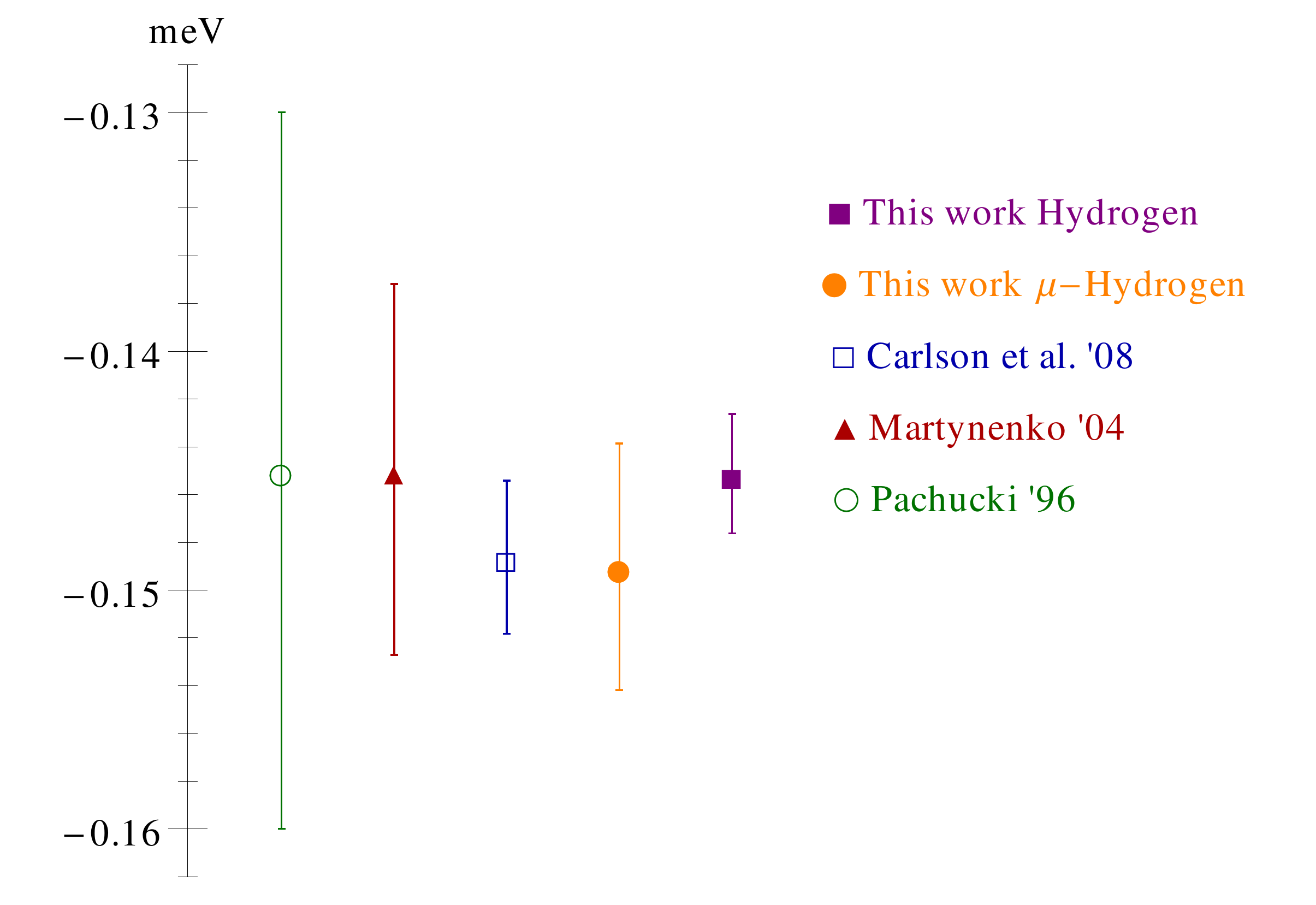}
\caption {\it Different predictions for $\delta \bar E^{\rm TPE}_{p \mu,\rm HF}(2S)$. For details see Table~\ref{tablecomparison} and the main text.}
\label{fig:comparison}
\end{center}
\end{figure}

We would like to finish comparing these numbers with other estimates using experimental information of the Sachs form factors and dispersion relations. We summarize the comparison in Table \ref{tablecomparison} and Fig.~\ref{fig:comparison}. Within uncertainties all numbers are consistent. The value of Ref.~\cite{Carlson:2008ke} is slightly lower, even though we get similar values for $c_{4,\rm TPE}^{p{\mu}}-c_{4,\rm TPE}^{p{e}}$. This can be traced back to the fact that they have a relatively big number (in absolute terms) for $\bar c_{4,\rm TPE}^{p{e}}$. They obtain $\bar c_{4,\rm TPE}^{p{e}}=-50.3(1.1)$. This number is slightly more than one sigma away from the number we obtain in Eq.~(\ref{c4peHF}) from the hydrogen. The value of Ref.~\cite{Martynenko:2004bt} uses the number for the polarizability term computed in Ref.~\cite{Faustov:2001pn}, and also incorporates the shift mentioned in Ref.~\cite{Carlson:2011af} (albeit this produces a very small change). 

\begin{table}[h]
\addtolength{\arraycolsep}{0.15cm}
\setlength\tabcolsep{7pt}
\begin{tabular}{|lc|c|c|c|c|c   l|}
 \hline 
& 
&\cite{pachucki1}& \cite{Martynenko:2004bt} & \cite{Carlson:2008ke} &{\rm Eq}.~(\ref{2Smuon}) &  {\rm Eq}.~(\ref{TPEpepmu}) & 
\\ \hline
&$\delta \bar E^{\rm TPE}_{p \mu,\rm HF}(2S)$
&$-0.145\,(15)$&$-0.1450\,(78)$&$-0.1486\,(32)$&$-0.1490\,(52)$ & $-0.1451\,(25)$&
\\ \hline
\end{tabular}
\caption{{\it Different predictions for $\delta \bar E^{\rm TPE}_{p \mu,\rm HF}(2S)$ in meV.  $\delta \bar E^{\rm TPE}_{p \mu,\rm HF}(1S)$ can trivially be obtained by multiplying them by $2^3$. The coefficient $\bar c_{4,\rm TPE}^{p{\mu}}$ can be obtained by multiplying them by the factor $\frac{2\pi m_p^2}{m_r^3 \alpha^5}=312.124\,\mathrm{meV}^{-1}$. The first three numbers refer to values obtained using experimental information of the Sachs form factors and dispersion relations. The fourth is from the measurement of the 2S hyperfine splitting in muonic hydrogen. The fifth is the prediction from chiral perturbation theory and the hydrogen hyperfine splitting.
}}
\label{tablecomparison}
\end{table}

If we focus strictly on the two-photon exchange contribution we should increase the error, as we expect the size of $\al K^{p\mu}$ to be of the order of $\al \times c_4^{p\mu}$. Therefore we add in quadrature a 1\% error to the error in Eq.  (\ref{TPEdifference}) and obtain $c_{4,TPE}^{p{\mu}}= -45.3(0.9)$. 
 
\section{Conclusions}

We have obtained a model-independent expression for the two-photon-exchange contribution to the hyperfine splitting in muonic hydrogen. 
We have used the relation of the Wilson coefficients of the spin-dependent dimension-six four-fermion operators of NRQED applied to the electron-proton and 
to the muon-proton sector. The difference can be reliably computed using chiral perturbation theory: 
\be
c_{4,\rm TPE}^{p\mu}-c_{4,\rm TPE}^{p e}=3.68(72)
\,.
\ee 
This allows us to give a 
precise model-independent determination of the Wilson coefficient of the spin-dependent dimension-six four-fermion operators of NRQED applied to the
muon-proton sector, and consequently of the two-photon exchange contribution to the hyperfine splitting in muonic hydrogen, 
which read 
\begin{eqnarray}
\bar c_{4,\rm TPE}^{p{\mu}}= -45.30(77) 
\;,
\qquad \delta \bar E^{\rm TPE}_{p \mu,\rm HF}
&=&\begin{cases}\displaystyle -1.161(20) &(1S),\\ -0.1451(25) &(2S)\,,
\end{cases}
\end{eqnarray}  
where we have used $\bar c_{4,\rm TPE}^{p{e}}=-48.69(3)$ from hydrogen. 

Together with the associated QED analysis, we obtain a prediction for the muonic hydrogen 
hyperfine that reads 
\begin{eqnarray}
E^{\rm th}_{p\mu,\rm HF}&=&\begin{cases}\displaystyle 182.623(27)\; {\rm meV} &{(1S)},\\ 
22.8123(33) \; {\rm meV}&{(2S)}\,,
\label{HFfinal}\end{cases}
\end{eqnarray} 
where the error is the linear sum of the errors of the two-photon exchange contribution in Eq.~(\ref{TPEpepmu}), and of the perturbative contribution in Eqs.~(\ref{1Smuonichydrogen}) and (\ref{2Smuonichydrogen}). 

A forthcoming PSI experiment is expected to reach a precision of the order of 100 MHz ($\sim 0.4$ $\mu$eV) 
for the hyperfine splitting of the muonic hydrogen \cite{Aldo}. Such precision may put into a test the universality of the lepton interactions. 

Finally, we also give a prediction for the following energy difference:
\be
8E_{p\mu,\rm HF}^{\rm th}(2S)-E_{p\mu,\rm HF}^{\rm th}(1S)=-0.124(3)\;{\rm meV}
\,.
\ee

\begin{acknowledgments}
This work was supported in part by the Spanish grants FPA2014-55613-P, FPA2013-
43425-P, and the Catalan grant SGR2014-1450.
\end{acknowledgments}

\end{document}